\documentclass{article}%
\usepackage[top=3cm, bottom=3cm, left=3cm, right=3cm]{geometry}
\usepackage{amsmath}
\usepackage{amsfonts}
\usepackage{amssymb}
\usepackage{graphicx}%
\providecommand{\U}[1]{\protect\rule{.1in}{.1in}}
\begin{document}

\title{QFT derivation of the decay law of an unstable particle with nonzero momentum}
\author{Francesco Giacosa\\\textit{Institute of Physics, Jan-Kochanowski University, }\\\textit{ul. Swietokrzyska 15, 25-406, Kielce, Poland.}\\\textit{Institute for Theoretical Physics, J. W. Goethe University, }\\\textit{ Max-von-Laue-Str. 1, 60438 Frankfurt, Germany.}}
\date{}
\maketitle

\begin{abstract}
We present a quantum field theoretical derivation of the nondecay probability
of an unstable particle with nonzero three-momentum $\mathbf{p}$. To this end,
we use the (fully resummed) propagator of the unstable particle, denoted as
$S,$ to obtain the energy probability distribution, called $d_{S}^{\mathbf{p}%
}(E)$, as the imaginary part of the propagator. The nondecay probability
amplitude of the particle $S$ with momentum $\mathbf{p}$ turns out to be, as
usual, its Fourier transform: $a_{S}^{\mathbf{p}}(t)=\int_{\sqrt{m_{th}%
^{2}+\mathbf{p}^{2}}}^{\infty}dEd_{S}^{\mathbf{p}}(E)e^{-iEt}$ ($m_{th}$ is
the lowest energy threshold in the energy frame, corresponding to the sum of
masses of the decay products). Upon a variable transformation, one can rewrite
it as $a_{S}^{\mathbf{p}}(t)=\int_{m_{th}}^{\infty}dmd_{S}^{\mathbf{0}%
}(m)e^{-i\sqrt{m_{th}^{2}+\mathbf{p}^{2}}t}$ [here, $d_{S}^{\mathbf{0}%
}(m)\equiv d_{S}(m)$ is the usual spectral function (or mass distribution) in
the rest frame]. Hence, the latter expression, previously obtained by
different approaches, is here confirmed in an independent and, most
importantly, covariant QFT-based approach. Its consequences are not yet fully
explored but appear to be quite surprising (such as the fact that usual
time-dilatation formula does not apply), thus its firm understanding and
investigation can be a fruitful subject of future research.

\end{abstract}

\section{Introduction}

The study of the decay law is a fundamental part of Quantum\ Mechanics (QM).
It is now theoretically
\cite{khalfin,misra,dega,ghirardi,facchiprl,ford,urbanolate,tunnelling,duecan}
and experimentally \cite{raizen,rothe} established that deviations from the
exponential decay exist, but they are usually small. Such deviations are also
present in\ Quantum\ Field Theory (QFT) \cite{duecan,zenoqft},

An interesting question addresses the decay of an unstable particle with
nonzero momentum $\mathbf{p.}$ In Refs.
\cite{khalfin2,shirokov,stefanovich,urbanowski,giacosaacta}, it was shown --by
using QM-based approaches enlarged to include special relativity-- that the
nondecay probability of an unstable particle $S$ with momentum $\mathbf{p}$ is
given by (in natural units)%
\begin{equation}
P_{S}^{\mathbf{p}}(t)=\left\vert a_{S}^{\mathbf{p}}(t)\right\vert ^{2}\text{
with }a_{S}^{\mathbf{p}}(t)=\int_{m_{th}}^{\infty}\mathrm{dm}d_{S}%
(m)e^{-i\sqrt{m^{2}+\mathbf{p}^{2}}t}\text{ ,} \label{1}%
\end{equation}
where $d_{S}(m)$ is the corresponding energy (or mass) distribution in its
rest frame [$\mathrm{dm}d_{S}(m)$ is the probability that the unstable state
has energy (or mass) between $m$ and $m+\mathrm{dm}$]. A review of the
derivation is presented in Sec. 2. Quite remarkably, there are many
interesting properties linked to this equation, which include deviations from
the standard dilatation formula, see below.

The purpose of this work is straightforward: we derive Eq. (1) in a QFT
framework (see Sec. 3).\ We thus confirm its validity and, as a consequence,
its peculiar consequences. We shall start from the relativistic propagator of
an unstable particle $S$. For definiteness, an underlying Lagrangian involving
scalar fields shall be introduced, but our discussion is valid for any
unstable field.

In this introduction, we recall some basic and striking features connected to
Eq. (1). The normalization%
\begin{equation}
\int_{m_{th}}^{\infty}\mathrm{dm}d_{S}(m)=1
\end{equation}
is a crucial feature of the spectral function, implying that $a_{S}%
^{\mathbf{p}}(0)=1$. It must be valid both in\ QM and in QFT. Here, without
loss of generality, we set the lower limit of the integral to $m_{th}\geq0$.
In fact, a minimal energy $m_{th}$ is present in each physical system; in
particular, for a (relevant for us) relativistic system, it is given by the
sum of the rest masses of the produced particles $(m_{th}=m_{1}+m_{2}%
+...\geq0).$ Clearly, for $\mathbf{p}=\mathbf{0}$,\ Eq. (1) reduces to the
usual expression
\begin{equation}
P_{S}^{rest}(t)=P_{S}^{\mathbf{0}}(t)=\left\vert a_{S}^{\mathbf{0}%
}(t)\right\vert ^{2}=\left\vert \int_{m_{th}}^{\infty}\mathrm{dm}%
d_{S}(m)e^{-imt}\right\vert ^{2}. \label{pndrest}%
\end{equation}
A detailed study of Eq. (1) shows peculiar features. First, the usual time
dilatation does not hold:
\begin{equation}
P_{S}^{\mathbf{p}}(t)\neq P_{S}^{rest}\left(  t\frac{M}{\sqrt{M^{2}%
+\mathbf{p}^{2}}}\right)  \text{ ,} \label{naive}%
\end{equation}
where $M$ is the mass of the state $S$ (defined, for instance, as the position
of the peak of the distribution $d_{S}(m);$ in general, however, other
definitions are possible, such as the real part of the pole of the propagator,
see Sec. 3. The point is that, no matter which definition one takes, the
expression (\ref{naive}) remains an inequality.)

In the exponential limit the spectral function of the state $S$ reads
\cite{ww,scully}
\begin{equation}
d_{S}^{BW}(m)=\frac{\Gamma}{2\pi}\left[  (m-M)^{2}+\Gamma^{2}/4\right]
^{-1}\text{,}%
\end{equation}
where $M$ is the `mass of the unstable state' corresponding to the peak. Even
if the spectral function $d_{S}^{BW}(m)$ is clearly unphysical because there
is no minimal energy ($m_{th}\rightarrow-\infty$), in many physical cases it
is a good approximation for a quite broad energy range . Here, the decay
amplitude and the decay law in the rest frame of the decaying particle
notoriously read%
\begin{equation}
a_{S}^{BW,\mathbf{0}}(t)=e^{-iMt-\Gamma t/2}\rightarrow P_{S}^{BW,rest}%
(t)=e^{-\Gamma t}.
\end{equation}
When a nonzero momentum is considered, the nondecay probability is still an
exponential given by%
\begin{equation}
P_{S}^{BW,\mathbf{p}}(t)=e^{-\Gamma_{\mathbf{p}}t}%
\end{equation}
where the width is \cite{giacosaacta}:
\begin{equation}
\Gamma_{\mathbf{p}}=\sqrt{2}\sqrt{\left[  \left(  M^{2}-\frac{\Gamma^{2}}%
{4}+\mathbf{p}^{2}\right)  ^{2}+M^{2}\Gamma^{2}\right]  ^{1/2}-\left(
M^{2}-\frac{\Gamma^{2}}{4}+\mathbf{p}^{2}\right)  }\text{ .} \label{gammap}%
\end{equation}
Clearly: $\Gamma_{\mathbf{p=0}}=\Gamma.$ One realizes that $\Gamma
_{\mathbf{p}}$ \textit{differs} from the naively expected standard
time-dilatation formula, according to which the decay width of an unstable
state with momentum $\mathbf{p}$ should simply be
\begin{equation}
\frac{\Gamma M}{\sqrt{\mathbf{p}^{2}+M^{2}}}=\gamma\Gamma\text{ .}
\label{einst}%
\end{equation}
Namely, the quantity $\gamma=\sqrt{\mathbf{p}^{2}+M^{2}}/M=1/\sqrt
{1-\mathbf{v}^{2}}$ is the usual dilatation factor for a state with (definite)
energy $M$. Deviations between Eq. (\ref{gammap}) and (\ref{einst}) are very
small, see the numerical discussion in\ Ref. \cite{giacosaacta}, and surely
not measurable by current experiments \cite{bailey}. Yet, the very fact that
deviations exist is very interesting and deserves further study.

It should be stressed that in this work we consider unstable states with a
definite momentum $\mathbf{p.}$ This is a subtle point: while for a state with
definite energy, a boost and a momentum translation are equivalent, this is
not so for an unstable state, since it is not an energy eigenstate. Even more
surprisingly, a boost of an unstable state is a quantum state whose nondecay
probability is actually zero: it is already decayed (on the contrary, its
survival probability present a peculiar time contraction \cite{giunti}). In
other words, a boosted muon consists of an electron and two neutrinos
\cite{stefanovich,giacosaacta}. In this sense, the boost mixes the Hilbert
subspace of the undecayed states with the subspace of the decay products, see
Ref. \cite{giacosaacta} for details. There, it is also discussed why the basis
of unstable states is contains states with definite three-momentum. Indeed,
the investigation of this paper also confirms this aspect: unstable states
with definite momentum naturally follow from the study of its propagator in QFT.

The paper is organized as follows: Sec. 2 we recall the QM derivation of Eq.
(1), while in Sec. 3 -- the key part of this paper-- we present this
derivation in a QFT context. In the end, in Sec. 4 we describe our conclusions.

\section{Recall of the QM-based derivation of Eq. (1)}

For completeness, we report here the \textquotedblleft
standard\textquotedblright\ derivation of Eq. (1). To this end, we use the
arguments presented in Ref. \cite{giacosaacta}, but similar ones can be found
in\ Refs. \cite{khalfin2,shirokov,stefanovich,urbanowski}.

We consider a system described by\emph{ }the Hamiltonian $H$, whose
eigenstates are denoted as%
\[
\left\vert m,\mathbf{p}\right\rangle =U_{\mathbf{p}}\left\vert m,\mathbf{0}%
\right\rangle \text{ ,}%
\]
where $U_{\mathbf{p}}$ is the unitary operator associated to the translation
in momentum space. Standard normalization expressions are assumed:
\begin{equation}
\left\langle m_{1},\mathbf{p}_{1}|m_{2},\mathbf{p}_{2}\right\rangle
=\delta(m_{1}-m_{2})\delta(\mathbf{p}_{1}-\mathbf{p}_{2})\text{ .}%
\end{equation}
The state $\left\vert m,\mathbf{p}\right\rangle $ has definite energy,
\begin{equation}
H\left\vert m,\mathbf{p}\right\rangle =\sqrt{\mathbf{p}^{2}+m^{2}}\left\vert
m,\mathbf{p}\right\rangle \text{ ,}%
\end{equation}
definite momentum,
\begin{equation}
\mathbf{P}\left\vert m,\mathbf{p}\right\rangle =\mathbf{p}\left\vert
m,\mathbf{p}\right\rangle ,
\end{equation}
as well as definite velocity $\mathbf{p}/\sqrt{\mathbf{p}^{2}+m^{2}}.$ Note,
assuming that the energy of $\left\vert m,\mathbf{p}\right\rangle $
is$\sqrt{\mathbf{p}^{2}+m^{2}},$ we have a relativistic spectrum.

Formally, the Hamiltonian can be written as
\begin{equation}
H=\int\mathrm{d}^{3}\mathrm{p}\int_{m_{th}}^{\infty}dm\sqrt{\mathbf{p}%
^{2}+m^{2}}\left\vert m,\mathbf{p}\right\rangle \left\langle m,\mathbf{p}%
\right\vert =\int\mathrm{d}^{3}\mathrm{p}H_{\mathbf{p}}%
\end{equation}
where%
\begin{equation}
H_{\mathbf{p}}=\int_{m_{th}}^{\infty}dm\sqrt{\mathbf{p}^{2}+m^{2}}\left\vert
m,\mathbf{p}\right\rangle \left\langle m,\mathbf{p}\right\vert \label{Hp}%
\end{equation}
is the effective Hamiltonian in the subspace of states with definite momentum
$\mathbf{p.}$

Let us now consider an unstable state $S$ in its rest frame. The corresponding
quantum state at rest is assumed to be%

\begin{equation}
\left\vert S,\mathbf{0}\right\rangle =\int_{m_{th}}^{\infty}\mathrm{dm}%
\alpha_{S}(m)\left\vert m,\mathbf{0}\right\rangle \text{ ,}%
\end{equation}
where $\alpha_{S}(m)$ is the probability amplitude that the state $S$ has
energy $m$. Hence, it is natural that the quantity $d_{S}(m)=\left\vert
\alpha_{S}(m)\right\vert ^{2}$ is the mass distribution: $d_{S}(m)\mathrm{dm}$
is the probability that the unstable particle $S$ has a mass between $m$ and
$m+\mathrm{dm}.$ As a consequence, $\int_{0}^{\infty}\mathrm{dm}d_{S}(m)=1$,
as already discussed in the introduction.

For the states of zero momentum, the Hamiltonian $H_{\mathbf{p}=\mathbf{0}}$
can be expressed in terms of the undecayed state $\left\vert S,\mathbf{0}%
\right\rangle $ and its decay products in the form of a Lee Hamiltonian
\cite{lee} (similar effective Hamiltonians are used also in quantum mechanics
\cite{ford,scully} and quantum field theory \cite{duecan,qcdeff}):%

\begin{align}
H_{\mathbf{p}=\mathbf{0}}  &  =\int_{m_{th}}^{\infty}\mathrm{dm}m\left\vert
m,\mathbf{0}\right\rangle \left\langle m,\mathbf{0}\right\vert \text{
}\nonumber\\
&  =M_{0}\left\vert S,\mathbf{0}\right\rangle \left\langle S,\mathbf{0}%
\right\vert +\int\mathrm{d}^{3}\mathrm{k}\omega(\mathbf{k})\left\vert
\mathbf{k},\mathbf{0}\right\rangle \left\langle \mathbf{k},\mathbf{0}%
\right\vert +\int\frac{\mathrm{d}^{3}\mathrm{k}}{(2\pi)^{3/2}}gf(\mathbf{k}%
)\left[  \left\vert S,\mathbf{0}\right\rangle \left\langle \mathbf{k}%
,\mathbf{0}\right\vert +\left\vert \mathbf{k},\mathbf{0}\right\rangle
\left\langle S,\mathbf{0}\right\vert \right]  \text{ ,} \label{H0}%
\end{align}
where $\left\vert \mathbf{k},\mathbf{0}\right\rangle $ represents a decay
product with vanishing total momentum: in the two-body decay case, $\left\vert
\mathbf{k},\mathbf{0}\right\rangle $ describes two particles, the first with
momentum $\mathbf{k}$ and the second with momentum $-\mathbf{k,}$ hence
\begin{equation}
\omega(\mathbf{k})=\sqrt{\mathbf{k}^{2}+m_{1}^{2}}+\sqrt{\mathbf{k}^{2}%
+m_{2}^{2}}\text{ .}%
\end{equation}
The last term in Eq. (\ref{H0}) represents the \textquotedblleft
mixing\textquotedblright\ between $\left\vert S,\mathbf{0}\right\rangle $ and
$\left\vert \mathbf{k},\mathbf{0}\right\rangle ,$ which cause the decay of the
former into the latter. Moreover, $g$ is a coupling constant and
$f(\mathbf{k)}$ encodes the dependence of the mixing on the momentum of the
produced particles. The explicit expressions connecting the states $\left\vert
\mathbf{k},\mathbf{0}\right\rangle $ to $\left\vert m,\mathbf{0}\right\rangle
$ formally reads%
\begin{equation}
\left\vert \mathbf{k},\mathbf{0}\right\rangle =\int_{m_{th}}^{\infty
}\mathrm{dm}\beta_{\mathbf{k}}(m)\left\vert m,\mathbf{0}\right\rangle
\label{k0}%
\end{equation}
where $\beta_{\mathbf{k}}(m)$ can be found by diagonalization of the
Hamiltonian (\ref{H0}).

Let us now consider an unstable state with definite momentum $\mathbf{p}$,
which is denoted as $\left\vert S,\mathbf{p}\right\rangle =U_{\mathbf{p}}$
$\left\vert S,\mathbf{0}\right\rangle $:%
\begin{equation}
\left\vert S,\mathbf{p}\right\rangle =\int_{m_{th}}^{\infty}\mathrm{dm}%
\alpha_{S}(m)\left\vert m,\mathbf{p}\right\rangle \text{ .} \label{sp}%
\end{equation}
The normalization%
\begin{equation}
\left\langle S,\mathbf{p}_{1}|S,\mathbf{p}_{2}\right\rangle =\delta
(\mathbf{p}_{1}-\mathbf{p}_{2})
\end{equation}
follows. Note, Eq. (\ref{sp}) is \textit{not} a state with definite velocity.
This is due to the fact that each state $\left\vert m,\mathbf{p}\right\rangle
$ in the superposition has a different velocity $\mathbf{p}/\sqrt
{\mathbf{p}^{2}+m^{2}}$. The subset of Hilbert space given by $\{\left\vert
S,\mathbf{p}\right\rangle \forall\mathbf{p\subset}R^{2}\}$ represents the set
of all undecayed quantum states of the system under study.

The form of the Hamiltonian $H_{\mathbf{p}}$ in term of the states $\left\vert
S,\mathbf{p}\right\rangle $ and $U_{\mathbf{p}}\left\vert \mathbf{k}%
,\mathbf{0}\right\rangle =\left\vert \mathbf{k},\mathbf{p}\right\rangle $ can
be in principle derived by using the expressions above. Together with Eq.
(\ref{sp}), one shall also take Eq. (\ref{k0}) and apply $U_{\mathbf{p}}$ in
order to get:
\begin{equation}
U_{\mathbf{p}}\left\vert \mathbf{k},\mathbf{0}\right\rangle =\left\vert
\mathbf{k},\mathbf{p}\right\rangle =\int_{m_{th}}^{\infty}\mathrm{dm}%
\beta_{\mathbf{k}}(m)\left\vert m,\mathbf{p}\right\rangle . \label{kp}%
\end{equation}
Then, once should invert Eqs. (\ref{sp}) and (\ref{kp}) and insert it into
$H_{\mathbf{p}}$ of Eq. (\ref{Hp}). However, its explicit expression is
definitely not trivial but, fortunately, also not needed in the present work.
Hence, we do not attempt to write it down here.

We now turn to the nondecay amplitudes. By starting from a properly normalized
state with zero momentum, $\left\vert S,\mathbf{0}\right\rangle /\sqrt
{\delta(\mathbf{0})},$ one obtains the usual expression%
\begin{align}
a_{S}^{\mathbf{0}}(t)  &  =\frac{1}{\delta(\mathbf{0})}\left\langle
S,\mathbf{0}\left\vert e^{-iHt}\right\vert S,\mathbf{0}\right\rangle =\frac
{1}{\delta(\mathbf{0})}\int_{m_{th}}^{\infty}\mathrm{dm}_{1}\mathrm{dm}%
_{2}\left\langle m_{1},\mathbf{0}\left\vert e^{-iHt}\right\vert m_{2}%
,\mathbf{0}\right\rangle \nonumber\\
&  =\int_{m_{th}}^{\infty}\mathrm{dm}d_{S}(m)e^{-imt}\text{ ,}%
\end{align}
in agreement with Eq. (\ref{pndrest}). The theory of decays is discussed in
great detail for the case $\mathbf{p}=\mathbf{0}$ in Refs.
\cite{ghirardi,ford,urbanolate,duecan} and refs. therein. Note, here the
nondecay probability coincides with the survival probability (that is, the
probability that the state did not change), but in general this is not the
case \cite{giacosaacta}.

Next, we consider a normalized unstable state $S$ with nonzero momentum:
$\left\vert S,\mathbf{p}\right\rangle /\sqrt{\delta(\mathbf{0})}.$ The
resulting non-decay probability amplitude
\begin{align}
a_{S}^{\mathbf{p}}(t)  &  =\frac{1}{\delta(\mathbf{0})}\left\langle
S,\mathbf{p}\left\vert e^{-iHt}\right\vert S,\mathbf{p}\right\rangle =\frac
{1}{\delta(\mathbf{0})}\int_{m_{th}}^{\infty}\mathrm{dm}_{1}\mathrm{dm}%
_{2}\left\langle m_{1},\mathbf{p}\left\vert e^{-iHt}\right\vert m_{2}%
,\mathbf{p}\right\rangle \nonumber\\
&  =\int_{m_{th}}^{\infty}\mathrm{dm}d_{S}(m)e^{-i\sqrt{m^{2}+\mathbf{p}^{2}%
}t}%
\end{align}
coincides with Eq. (1), hence concluding our derivation.

In principle, one could also start from the Hamiltonian $H_{\mathbf{p}}$ and
obtain the energy distribution associated to this state, denoted as
$d_{S}^{\mathbf{p}}(E).$ Then, $a_{S}^{\mathbf{p}}(t)$ should also emerge as
the Fourier transform of the latter. This is hard to do here, since the
explicit expression of $H_{\mathbf{p}}$ in terms of $\left\vert S,\mathbf{p}%
\right\rangle $ and $\left\vert \mathbf{k},\mathbf{p}\right\rangle $ was not
written down (this is not an easy task). Quite interestingly, in the framework
of QFT the function $d_{S}^{\mathbf{p}}(E)$ can be easily determined, see Sec. 3.

As a last comment of this section, we recall that the general nondecay
probability of an arbitrary state $\left\vert \Psi\right\rangle $ reads:%
\begin{equation}
P_{\left\vert \Psi\right\rangle }(t)=\int\mathrm{d}^{3}\mathrm{p}\left\vert
\left\langle S,\mathbf{p}\left\vert e^{-iHt}\right\vert \Psi\right\rangle
\right\vert ^{2}\text{ ,}%
\end{equation}
whose interpretation is straightforward: we project $\left\vert \Psi
\right\rangle $ onto the basis of undecayed states. In general, $P_{\left\vert
\Psi\right\rangle }(0)$ is not unity. Notice also that $P_{\left\vert
\Psi\right\rangle }(t)$ is \textit{not} the survival probability of the state
$\left\vert \Psi\right\rangle $ (a state can change with time, but still be
undecayed if it is a different superposition of $\left\vert S,\mathbf{p}%
\right\rangle $).

When a boost $U_{\mathbf{v}}$ on the state with zero momentum (and hence with
zero velocity) $\left\vert S,\mathbf{0}\right\rangle $ is considered, the
resulting state reads \cite{giacosaacta}:
\begin{equation}
\left\vert \varphi_{\mathbf{v}}\right\rangle =U_{\mathbf{v}}\left\vert
S,\mathbf{0}\right\rangle =\int_{m_{th}}^{\infty}\mathrm{dm}\alpha_{S}%
(m)\sqrt{m}\gamma^{3/2}\left\vert m,m\gamma\mathbf{v}\right\rangle \text{ ,}
\label{phiv}%
\end{equation}
where $\gamma=(1-\mathbf{v}^{2})^{-1/2}.$ In fact, each element of the
superposition, $\left\vert m,m\gamma\mathbf{v}\right\rangle $, has velocity
$\mathbf{v}$. Of course, $\left\vert \varphi_{\mathbf{v}}\right\rangle $ is
not an eigenstate of momentum, since each element in Eq. (\ref{phiv}) has a
different momentum $\mathbf{p}=m\gamma\mathbf{v}$\textbf{.} In this respect
the state $\left\vert S,\mathbf{0}\right\rangle $ is special: it is the only
state which has at the same time definite momentum and definite velocity (both
of them vanishing). As mentioned in the Introduction, the nondecay probability
associated to $\left\vert \varphi_{\mathbf{v}}\right\rangle $ vanishes:%
\begin{equation}
P_{\left\vert \varphi_{\mathbf{v}}\right\rangle }(t)=0\text{ }\forall
\mathbf{v\neq0}.
\end{equation}
As soon as a nonzero velocity is considered, the state has decayed. This
result is quite surprising but also rather `delicate': when a wave packet is
considered, $P_{nd}^{\left\vert \varphi_{\mathbf{v}}\right\rangle }(t)$ is
nonzero (even if it is not $1$ for $t=0$) \cite{giacosaacta}.

\section{Covariant QFT derivation of Eq. (1)}

Let us consider an unstable particle described by the field $S(x)\equiv
S(t,\mathbf{x})$. For simplicity one can take a scalar field $S$ with bare
mass $M_{0}$ coupled to two scalar fields $\varphi_{1}$ (with mass $m_{1}$)
and $\varphi_{2}$ (with mass $m_{2}$) via the interaction term $gS\varphi
_{1}\varphi_{2}$, leading to the QFT Lagrangian%
\begin{equation}
\mathcal{L}=\frac{1}{2}\left[  \left(  \partial_{\mu}\varphi_{1}\right)
^{2}-m_{1}^{2}\varphi_{1}^{2}\right]  +\frac{1}{2}\left[  \left(
\partial_{\mu}\varphi_{2}\right)  ^{2}-m_{2}^{2}\varphi_{2}^{2}\right]
+\frac{1}{2}\left[  \left(  \partial_{\mu}S\right)  ^{2}-M_{0}^{2}%
S^{2}\right]  +gS\varphi_{1}\varphi_{2}\text{ .}\label{lag}%
\end{equation}
This is the QFT counterpart of the previous section. However, our discussion
is in no way limited to this scalar theory.

The (full) propagator of the state $S$ (details in\ Ref. \cite{lupo}) reads:%
\begin{equation}
\Delta_{S}(p^{2})=\frac{1}{p^{2}-M_{0}^{2}+\Pi(p^{2})+i\varepsilon}\text{ with
}p^{2}=E^{2}-\mathbf{p}^{2}\text{ ,} \label{prop}%
\end{equation}
where $E=p^{0}$ is the energy and $\mathbf{p}$ the three-momentum. Because of
covariance, $\Delta_{S}(p^{2})$ depends only on $p^{2}$. The quantity
$\Pi(p^{2})$ is the one-particle irreducible diagram. Its calculation is of
course non trivial (it requires a proper regularization), but it is not needed
for our purposes. The imaginary part
\begin{equation}
\operatorname{Im}\Pi(p^{2})=\sqrt{p^{2}}\Gamma(\sqrt{p^{2}})=\frac{\left\vert
\mathbf{k}\right\vert }{8\pi\sqrt{p^{2}}}g^{2}f_{\Lambda}^{2}(\left\vert
\mathbf{k}\right\vert )+...,
\end{equation}
where dots refer to higher orders, which are however typically very small
\cite{schneitzer}. Once $\operatorname{Im}\Pi(p^{2})$ is fixed,
$\operatorname{Re}\Pi(p^{2})$ can be determined by dispersion relations (for
an example of this technique, see e.g. Ref. \cite{a0}). The quantity
$\Gamma^{tl}=\Gamma(\sqrt{p^{2}}=M)$, is the usual tree-level decay width,
hence in the exponential limit the decay law $P_{S}(t)=e^{-\Gamma^{tl}t}$ must
be reobtained. As mentioned in the Introduction, an unstable state has not a
definite mass: this is why different definitions for $M$ (which is not the
bare mass $M_{0}$ entering in Eq. ()) are possible: $\operatorname{Re}%
\Delta_{S}^{-1}(p^{2}=M^{2})=0$ (zero of the real part of the denominator), or
$\operatorname{Re}\left[  \sqrt{s_{pole}}\right]  ,$ with $\Delta_{S}%
^{-1}(s_{pole})=0$ (real part of the pole), or the maximum of the spectral
function defined below.

We also recall that
\begin{equation}
\left\vert \mathbf{k}\right\vert =\sqrt{\frac{p^{4}+(m_{1}^{2}-m_{2}^{2}%
)^{2}-2p^{2}(m_{1}^{2}+m_{2}^{2})}{4p^{2}}}%
\end{equation}
coincides, for the on-shell decay, with the the three-momentum of one of the
outgoing particles. The vertex function $f_{\Lambda}(\left\vert \mathbf{k}%
\right\vert )$ is a proper regularization which fulfills the condition
$f_{\Lambda}(\left\vert \mathbf{k}\right\vert \rightarrow0)=1$ and describes
the high-energy behavior of the theory (its UV completion), hence the
parameter $\Lambda$ is some (very) high energy scale; $f_{\Lambda}(\left\vert
\mathbf{k}\right\vert )$ is formally not present in Eq. (\ref{lag}) since it
appears in the regularization procedure, but it can be included directly in
the Lagrangian by rendering it nonlocal \cite{nl} in a way that fulfills
covariance \cite{covariant}. In a renormalizable theory (such as the one of
Eq. (\ref{lag})), the dependence on $\Lambda$ disappears in the low-energy limit.

The properties outlined above, although in general very important in specific
calculations, turn out to be actually secondary to the proof that we present
below, where only the formal expression of the propagator of Eq. (\ref{prop})
is relevant. Moreover, even when the unstable particle is not a scalar, one
can always define a scalar part of the propagator which looks just as in\ Eq.
(\ref{prop}), then the outlined properties apply, \textit{mutatis mutandis},
to each QFT Lagrangian.

As a next step, upon introducing the Mandelstam variable $s=p^{2}$, the
function $F(s)$ defined as
\begin{equation}
F(s)=\frac{1}{\pi}\operatorname{Im}\left[  \Delta_{S}(p^{2}=s)\right]
\end{equation}
fulfills the normalization condition:%

\begin{equation}
\int_{s_{th}}^{\infty}\mathrm{ds}F(s)=1\text{ ,} \label{normF}%
\end{equation}
where $s_{th}=m_{th}^{2}$ is the minimal squared energy. For the case of Eq.
(\ref{lag}), one has obviously $s_{th}=\left(  m_{1}+m_{2}\right)  ^{2}.$ The
normalization is a consequence of the K\"{a}ll\'{e}n--Lehmann representation
\cite{peskin}%
\begin{equation}
\Delta_{S}(p^{2})=\int_{s_{th}}^{\infty}\mathrm{ds}\frac{F(s)}{p^{2}%
-s+i\varepsilon}\text{ ,}%
\end{equation}
in which the propagator $\Delta_{S}(p^{2})$ has been rewritten as the `sum' of
free propagators $\left[  p^{2}-s+i\varepsilon\right]  ^{-1},$ each one of
them weighted by $F(s)$: $\mathrm{ds}F(s)$ is the probability that the the
squared mass lies between $s$ and $s+\mathrm{ds}.$ Of course, the
normalization (\ref{normF}) is a central feature. For the detailed proof of
it, we refer to Ref. \cite{lupoprd}. Here we recall a simple version of it,
which is obtained by assuming the rather strong requirement $\Pi(p^{2})=0$ for
$p^{2}>\Lambda^{2},$ where $\Lambda$ is a high-energy scale (no matter how
large). Under this assumption%
\begin{equation}
\Delta_{S}(p^{2})=\frac{1}{p^{2}-M_{0}^{2}+\Pi(p^{2})+i\varepsilon}%
=\int_{s_{th}}^{\Lambda^{2}}\mathrm{ds}\frac{F(s)}{p^{2}-s+i\varepsilon}\text{
.}%
\end{equation}
Then, upon taking a certain value $p^{2}\gg\Lambda^{2},$ the previous equation
reduces to%
\begin{equation}
\frac{1}{p^{2}}=\int_{s_{th}}^{\Lambda^{2}}\mathrm{ds}\frac{F(s)}{p^{2}%
}\mathrm{\rightarrow}\int_{s_{th}}^{\Lambda^{2}}\mathrm{ds}F(s)=1.
\end{equation}
The general case in which $\Pi(p^{2}\rightarrow\infty)=0$ requires more steps,
but the final result of Eq. (\ref{normF}) still holds \cite{lupoprd}.

Let us now consider the rest frame of the decaying particle: $\mathbf{p}%
=\mathbf{0}$, $s=p^{2}=E^{2}=m^{2}$. Here, upon a simple variable change
$(m=\sqrt{s}$), we obtain the mass distribution (or spectral function)
$d_{S}^{\mathbf{p}=\mathbf{0}}(m)$ through the equation%

\begin{equation}
\mathrm{dm}d_{S}^{\mathbf{p}=\mathbf{0}}(m)=\mathrm{ds}F(s)\text{ ,}%
\end{equation}
out of which:%
\begin{equation}
d_{S}(m)=d_{S}^{\mathbf{p}=\mathbf{0}}(m)=2mF(s=m^{2})\text{ . }%
\end{equation}
As already mentioned, $\mathrm{dm}d_{S}(m)$ is the probability that the
particle $S$ has a mass between $m$ and $m+\mathrm{dm}$ \cite{lupo,salam}. In
this context, the normalization
\begin{equation}
\int_{m_{th}}^{\infty}\mathrm{dm}d_{S}(m)=1
\end{equation}
follows from Eq. (\ref{normF}). Once the function $d_{S}(m)$ is identified as
the mass distribution of the undecayed quantum state, the non-decay
probability's amplitude $a_{S}^{\mathbf{0}}(t)$ can be obtained by repeating
the steps of Sec. 2. The result coincides, as expected, with Eq.
(\ref{pndrest}). Yet, it should be stressed that the unstable quantum state
$\left\vert S,\mathbf{0}\right\rangle $ characterized by the distribution
$d_{S}(m)$ is not simply given by $a_{\mathbf{0}}^{\dagger}\left\vert
0_{PT}\right\rangle ,$ where $\left\vert 0_{PT}\right\rangle $ is the
perturbative vacuum and $a_{\mathbf{p}}^{\dagger}$ the creator operator of the
non-interacting field $S$. The case of neutrino oscillations shows a similar
situation: the state corresponding to a certain flavour, such as the neutrino
$\nu_{e}$, must be constructed with due care by making use of Bogolyubov
transformations \cite{blasone}. Along this line, the exact and formal
determination of the state $\left\vert S,\mathbf{0}\right\rangle ,$
corresponding to the mass distribution $d_{S}(m)$, in the context of QFT
requires a generalization of Bogolyubov transformations and is not an easy
task (it is left for the future). Nevertheless, it is not needed for the
purpose of this paper.

Let us now consider the particle $S$ moving with a certain momentum
$\mathbf{p}$. Upon using $s=E^{2}-\mathbf{p}^{2},$ the energy distribution -as
function of $E$- is obtained by%
\begin{equation}
\mathrm{dE}d_{S}^{\mathbf{p}}(E)=\mathrm{ds}F(s)\text{ ,}%
\end{equation}
leading to%

\begin{equation}
d_{S}^{\mathbf{p}}(E)=2EF(s=E^{2}-\mathbf{p}^{2})=\frac{E}{\sqrt
{E^{2}-\mathbf{p}^{2}}}d_{S}(\sqrt{E^{2}-\mathbf{p}^{2}})\text{ .} \label{dsp}%
\end{equation}
The quantity $\mathrm{dE}d_{S}^{\mathbf{p}}(E)$ is the probability that the
particle $S$ with definite momentum $\mathbf{p}$ has an energy between $E$ and
$E+\mathrm{dE}$ (clearly, $d_{S}^{\mathbf{p=0}}(E)=d_{S}(m=E)$). Also in this
case, the normalization%
\begin{equation}
\int_{\sqrt{m_{th}^{2}+\mathbf{p}^{2}}}^{\infty}\mathrm{dE}d_{S}^{\mathbf{p}%
}(E)=1
\end{equation}
is a consequence of Eq. (\ref{normF}). When $d_{S}(m)$ has a maximum at $M$,
then $d_{S}^{\mathbf{p}}(E)$ has a maximum at $\sim\sqrt{M^{2}+\mathbf{p}^{2}%
}$. Note, the very fact that the propagator depends on $p^{2}=E^{2}%
-\mathbf{p}^{2}$ allows to determine the spectral function $d_{S}^{\mathbf{p}%
}(E)$ for a definite momentum $\mathbf{p,}$ that corresponds to the state
$\left\vert S,\mathbf{p}\right\rangle $ of Sec. 2.

The nondecay probability's amplitude for a state $S$ moving with momentum
$\mathbf{p}$ is then given by%

\begin{equation}
a_{S}^{\mathbf{p}}(t)=\int_{\sqrt{m_{th}^{2}+\mathbf{p}^{2}}}^{\infty
}\mathrm{dE}d_{S}^{\mathbf{p}}(E)e^{-iEt}\text{ ,}%
\end{equation}
where we have taken into account that the minimal energy is given by
$\sqrt{m_{th}^{2}+\mathbf{p}^{2}}$.

This expression can be manipulated by using Eq. (\ref{dsp}) and via a change
of variable:
\begin{align}
a_{S}^{\mathbf{p}}(t)  &  =\int_{\sqrt{m_{th}^{2}+\mathbf{p}^{2}}}^{\infty
}\mathrm{dE}d_{S}^{\mathbf{p}}(E)e^{-iEt}=\int_{\sqrt{m_{th}^{2}%
+\mathbf{p}^{2}}}^{\infty}\mathrm{dE}\frac{E}{\sqrt{E^{2}-\mathbf{p}^{2}}%
}d_{S}(\sqrt{E^{2}-\mathbf{p}^{2}})e^{-iEt}\nonumber\\
&  =\int_{m_{th}}^{\infty}\mathrm{dm}d_{S}(m)e^{-i\sqrt{m^{2}+\mathbf{p}^{2}%
}t}\text{ ,}%
\end{align}
which coincides \textit{exactly} with Eq. (\ref{1}), as we wanted to
demonstrate. Thus, we confirm the validity of Eq. (1) in a covariant QFT-based framework.

\section{Conclusions}

The decay law of moving unstable particles is an interesting subject that
connects special relativity to QM and QFT. An important aspect is the validity
of Eq. (1), which expresses the nondecay probability of a state with nonzero
momentum and whose standard derivation is reviewed in Sec. 2.

The main contribution of this paper has been the derivation of a quantum field
theoretical proof of Eq. (1).\ To this end, we started from the (scalar part
of the) propagator of an unstable quantum field, denoted as $S$. Then, we have
determined the energy distribution of the state $S$ with definite momentum
$\mathbf{p}$, out of which the survival's probability amplitude is calculated.

As discussed in the Introduction, there are interesting and peculiar
consequences of Eq. (1). Future studies are definitely needed to further
understand the properties of a decay of a moving unstable particle and to look
for feasible experimental tests.

\bigskip

\textbf{Acknowledgments: }I thank S. Mr\'{o}wczy\'{n}ski and G. Pagliara for
useful discussions.

\end{document}